\documentclass[5p,twocolumn,times,number]{elsarticle}
\usepackage{amsmath}
\usepackage{amssymb}
\usepackage{graphics}
\usepackage{graphicx}
\usepackage{epsfig}
\usepackage{dcolumn}
\usepackage{bm}
\usepackage{lineno}

\begin{document}
\begin{frontmatter}

\title{Study of gain variation as a function of physical parameters of GEM foil}
\author{Supriya~Das\corref{cor} on behalf of the ALICE Collaboration}
\ead{Supriya.Das@cern.ch}
\address{Department of Physics, Bose Institute, Kolkata - 700091, India.}
\date{today}

\cortext[cor]{Corresponding author}

%\maketitle
\begin{abstract}
The ALICE experiment at LHC has planned to upgrade the TPC by replacing the MWPC with GEM based 
detecting elements to 
restrict the IBF to a tolerable value. However the variation of the gain as a function of physical 
parameters of industrially produced large size GEM foils is needed to be studied as a part of the 
QA procedure for the detector. The size of the electron avalanche and consequently the gain for GEM based detectors depend on the 
electric field distribution inside the holes. Geometry of a hole plays an important role in defining the 
electric field inside it. In this work we have studied the variation of the gain as a function of 
the hole diameters using Garfield++ simulation package.
\end{abstract}


\end{frontmatter}

\section{Introduction}

A Large Ion Collider Experiment (ALICE) uses a Time Projection Chamber (TPC) \cite{TPC04} as 
the central tracking detector for charged particle tracking. Currently the readout rate of the TPC is 
limited to 3.5 kHz due to the use of triggered Gating Grid (GG).
To cope up with the expected event rate of 50 kHz for Pb-Pb collisions in the LHC RUN 3 after
the Long Shutdown 2 (LS2), a major upgrade program for the ALICE TPC has been taken up \cite{Gunji14}.
During this upgrade, the Multi Wire Proportional Chambers (MWPC) at the end cap of the TPC will be 
replaced with detecting elements based on Gas Electron Multiplier (GEM) technology to restrict the 
Ion Back Flow (IBF) into the drift volume to a tolerable limit in the absence of the GG. 

The basic working principle of GEM detector
is the electron multiplication that happens when primary electrons pass through small holes in the
GEM foils where a very high electric field is imposed \cite{Sauli}. However, the electric field
distribution inside the holes depends  on the shape and size of the holes as well as the thickness 
of the polyimide foil. In consequence, the local variation of these parameters over the entire 
area of the foil results in variation of the gain. Due to the complications in the 
fabrication processes of GEM foils, deviations from the design parameters over the entire area of the 
foils are not rare, specially when large sized foils are fabricated. Such variation of gain 
as a function of the hole diameter has been reported in a recent work \cite{Timo15}. But it is not 
possible to measure the gain for each individual foil before they are used to fabricate 
the detectors, when a large number of foils are involved. So an independent evaluation of the 
variation of gain as a function of the physical parameters from simulation is necessary for better 
operation of the detector.

We have used a widely used commercially available finite element tool, ANSYS \cite{Ansys} to create the 
electric field distribution inside the detector and open source Garfield++ \cite{Garfield} package to 
simulate the electron avalanche through the holes. A more detailed description of the 
methodologies and results are discussed in section~\ref{sim} and section~\ref{res} respectively.  

\vspace{-0.4cm}
\section{Simulation of electron avalanche}\label{sim}

\begin{figure}[htb!]
\begin{center}
\includegraphics[scale=0.3]{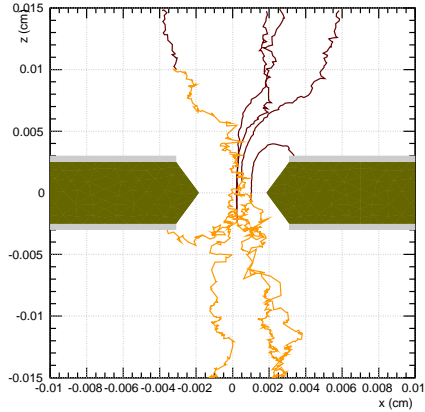}
\caption{\label{fig_avalanche}Electron multiplication across a typical double mask GEM hole.
Both electrons (yellow) and ions (brown) are shown.}
\end{center}
\end{figure}

\begin{figure}[htb!]
\begin{center}
\includegraphics[scale=0.34]{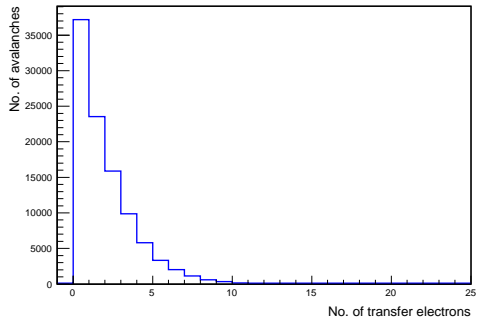}
\caption{\label{fig_electron_distribution}Distribution of the number of transfer electrons for a typical
hole configuration of 70-50-70 $\mu$m for top-middle-bottom hole diameter of a double mask GEM.}
\end{center}
\end{figure}

The detecting element of GEM detector is a 50 $\mu$m polyimide foil sandwiched between 5 $\mu$m
layers of copper on both sides. For foils produced with double mask technology, the holes have 
an hour-glass like shape where the diameter of
the outer openings have typical values of 70 $\mu$m and the diameter of the constricted part is
50 $\mu$m. Single mask technology produces holes with conical shape with typical hole 
diameters of 70 $\mu$m and 60 $\mu$m on either sides of the cone. A high electric field
inside the hole is created and controlled by applying voltages
on the metal layers. Electron multiplication happens when the electric field in the hole reaches 
a typical value of 10 kV/cm which is the limit of avalanche formation for the gas. The geometry
of the hole, material properties of the GEM, the voltage applied to the metal layers and the 
boundary conditions are defined in ANSYS, which calculates the electric field mesh using finite
element technique. The overall formation of the electron avalanche (Fig.~\ref{fig_avalanche}) is 
simulated using Garfield++.
This program has interfaces with HEED \cite{Heed}, which calculates the energy loss and ionisation
of the gas molecules by moving charged particles and MAGBOLTZ\cite{Magboltz}, which computes the transport 
properties of the electrons subjected to the electric field. Finally for every single primary electron,
the number of secondaries that cross the hole and reach certain distance below the bottom of 
the foil is counted. A distribution of this number for many thousands of avalanches is shown in 
Fig.~\ref{fig_electron_distribution}. The mean of this distribution gives the gain.

%%%%%%%%%%%%%%%%%%%%%%%%%%%%%%%%%%%%%%%%%%%%%%%%%%%%%%%%%%%%%%%%%%%

%%%%%%%%%%%%%%%%%%%%%%%%%%%%%%%%%%%%%%%%%%%%%%%%%%%%%%%%%%%%%%%%%%%

%%%%%%%%%%%%%%%%%%%%%%%%%%%%%%%%%%%%%%%%%%%%%%%%%%%%%%%%%%%%%%%%%%%

%%%%%%%%%%%%%%%%%%%%%%%%%%%%%%%%%%%%%%%%%%%%%%%%%%%%%%%%%%%%%%%%%%%

\vspace{-0.4cm}
\section{Results}\label{res}

\begin{figure}[htb!]
\begin{center}
\includegraphics[scale=0.345]{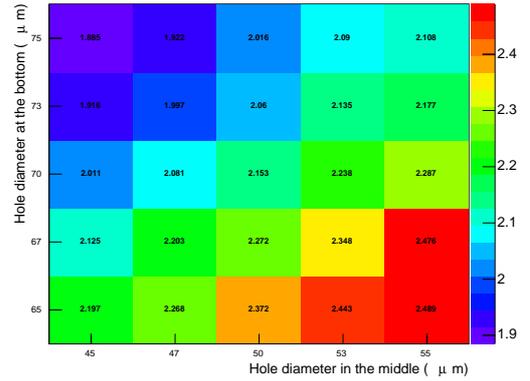}
\caption{\label{fig_2d_gain}Gain as a function of the hole diameters at the bottom and
in the middle, keeping the diameter at the top fixed at 70 $\mu$m.}
\end{center}
\end{figure}

\begin{figure}[htb!]
\begin{center}
\includegraphics[scale=0.345]{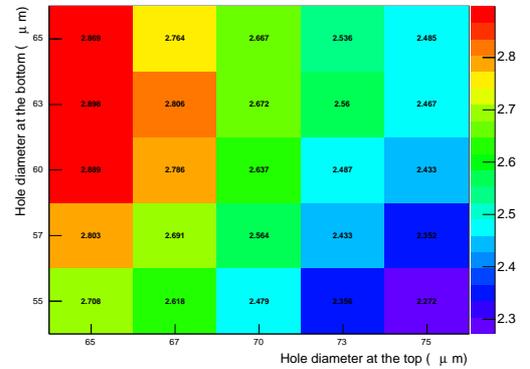}
\caption{\label{fig_2d_gain_single_mask}Gain as a function of the hole diameters at 
the bottom and that at the top for single mask GEM foil.}
\end{center}
\end{figure}

It is important to mention here that the charging up effect of the GEM foils has not been considered in 
this study. We have taken five values of each of the three hole diameters to calculate the gain for 
bi-conical holes, on the other hand five values of each of the two hole diameters have been taken for 
conical holes. 100 k avalanches have been simulated for each setting. $\text{Ar+CO}_\text{2}$ (70:30) gas 
mixture has been used for all the calculations. The values of the drift field ($E_d$)  and those of the 
induction field ($E_i$) have been taken as 400 V/cm and 1 kV/cm respectively. A potential difference 
($\Delta V_{GEM}$) of 267 V has been applied across the foil. The variation of the gain as a function 
of different hole diameters for the double and the single mask GEM foils are shown in 
Fig.~\ref{fig_2d_gain} and Fig.~\ref{fig_2d_gain_single_mask} respectively. 

%%%%%%%%%%%%%%%%%%%%%%%%%%%%%%%%%%%%%%%%%%%%%%%%%%%%%%%%%%%%%%%%%%%

%%%%%%%%%%%%%%%%%%%%%%%%%%%%%%%%%%%%%%%%%%%%%%%%%%%%%%%%%%%%%%%%%%%

%%%%%%%%%%%%%%%%%%%%%%%%%%%%%%%%%%%%%%%%%%%%%%%%%%%%%%%%%%%%%%%%%%%

\vspace{-0.3cm}
\section{Summary and outlook}

We have studied the variation of the gain of single GEM detector as a function of the shape of the holes using 
Garfield++ simulation package. The results reveal that this variation may reach up to 17-18\% 
for a variation of 5 $\mu$m in the hole diameter. It is understood that the charging up 
of the GEM foils in course of time might have effect on the results. However these findings are 
needed to be compared with the measured data for similar variation. The gain variation will be 
studied by changing other 
parameters such as gas mixture and $\Delta V_{GEM}$ as a continuation of this work. The results
of this study will help in QA of the GEM foils where a large number of them is required such as in 
ALICE TPC.

\vspace{-0.4cm}
\section*{Acknowledgements}
%\newline

This work was partially supported by the research grant SR/MF/PS-01/2014-BI from Department of 
Science and Technology, Govt. of India.

\end{document}